\documentstyle[epsf,12pt]{article}

\newcommand{\DMTH}{\mbox{$\Delta M(T,H)$}}

\newcommand{\DM}{\mbox{$\Delta M$}}

\newcommand{\wEnk}{\mbox{$W(E_{nk_z})$}}

\newcommand{\wEnkPAW}{\mbox{$W_{\rm PAW}(E_{nk_z})$}}

\newcommand{\eg}{{\it e.g.}}

\newcommand{\etal}{{\it et al.}}

\newcommand{\lsim}{\stackrel{<}{_\sim}}

\newcommand{\gsim}{\stackrel{>}{_\sim}}

\newcommand{\reffig}[1]{~\ref{#1}}

\newcommand{\figi}{
\epsfxsize=0.95\textwidth
\begin{figure}[h]
\epsfbox{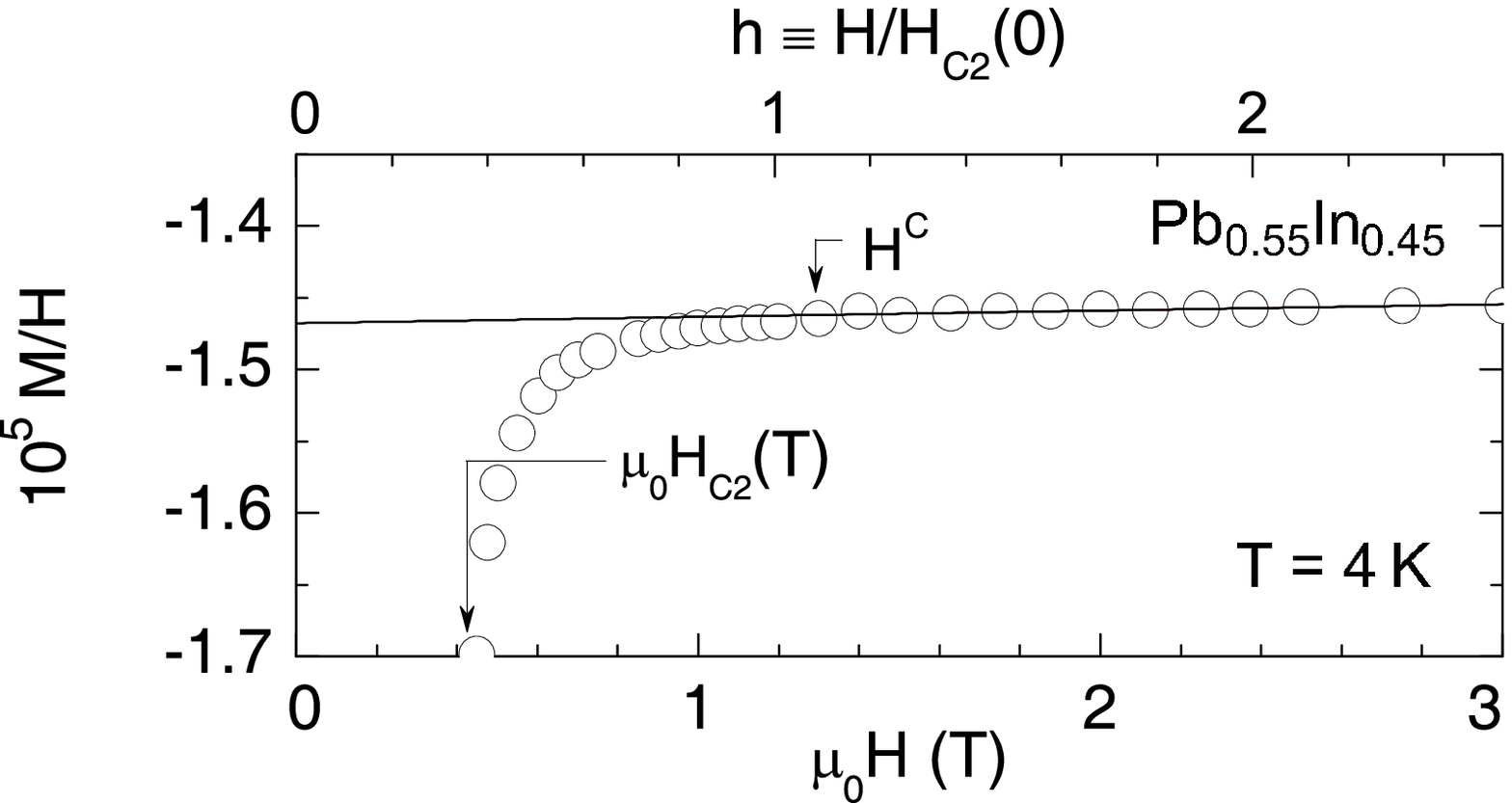}
\caption{\label{Fig1}An example, corresponding to the Pb$_{0.55}$In$_{0.45}$ alloy, of the as-measured magnetic susceptibility versus applied magnetic field at constant temperature below $T_{C0}$. The solid line is the background susceptibility, obtained by a linear fit in a region far from the superconducting transition, in this example from 2~Tesla to 3~Tesla [i.e., from $\sim 5H_{C2}(T)$ to $\sim 7H_{C2}(T)$]. } 
\end{figure}}

\newcommand{\figii}{
\begin{figure}[t]
\epsfxsize=0.95\textwidth
\epsfbox{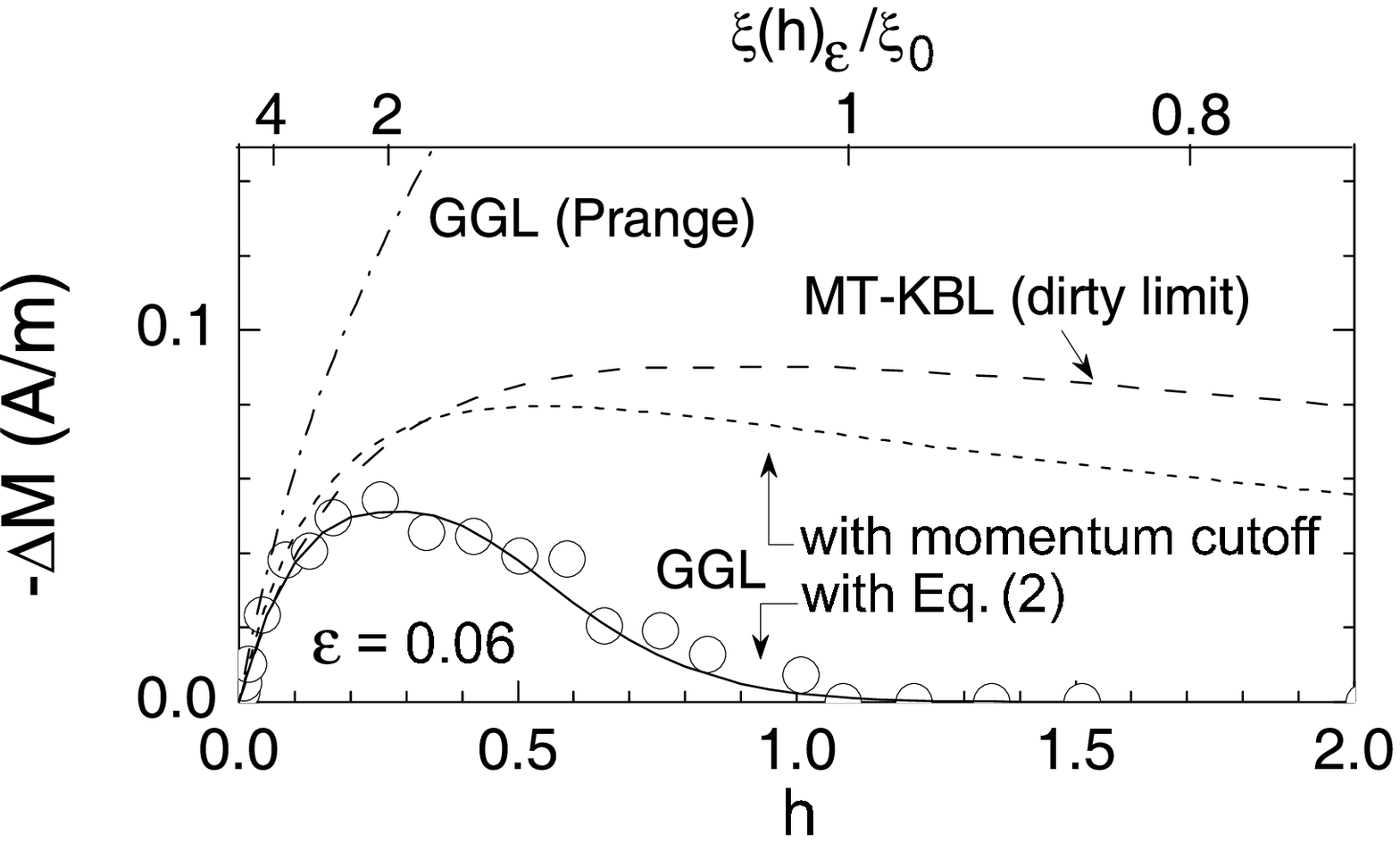}
\caption{\label{Fig2}An example, corresponding to the Pb$_{0.55}$In$_{0.45}$ alloy, of the reduced-magnetic-field dependence of the fluctuation-induced magnetization, at $\varepsilon=0.06$. The upper scale shows that for $h\gsim0.2$, $\xi(h)_\varepsilon\lsim2\xi_0$ and it already suggests that in this high-field region the SF behaviour is dominated by the uncertainity principle. The curves correspond to different theoretical approches, as explained in the main text. }
\end{figure}}

\newcommand{\figiii}{
\begin{figure}[t]
\epsfbox{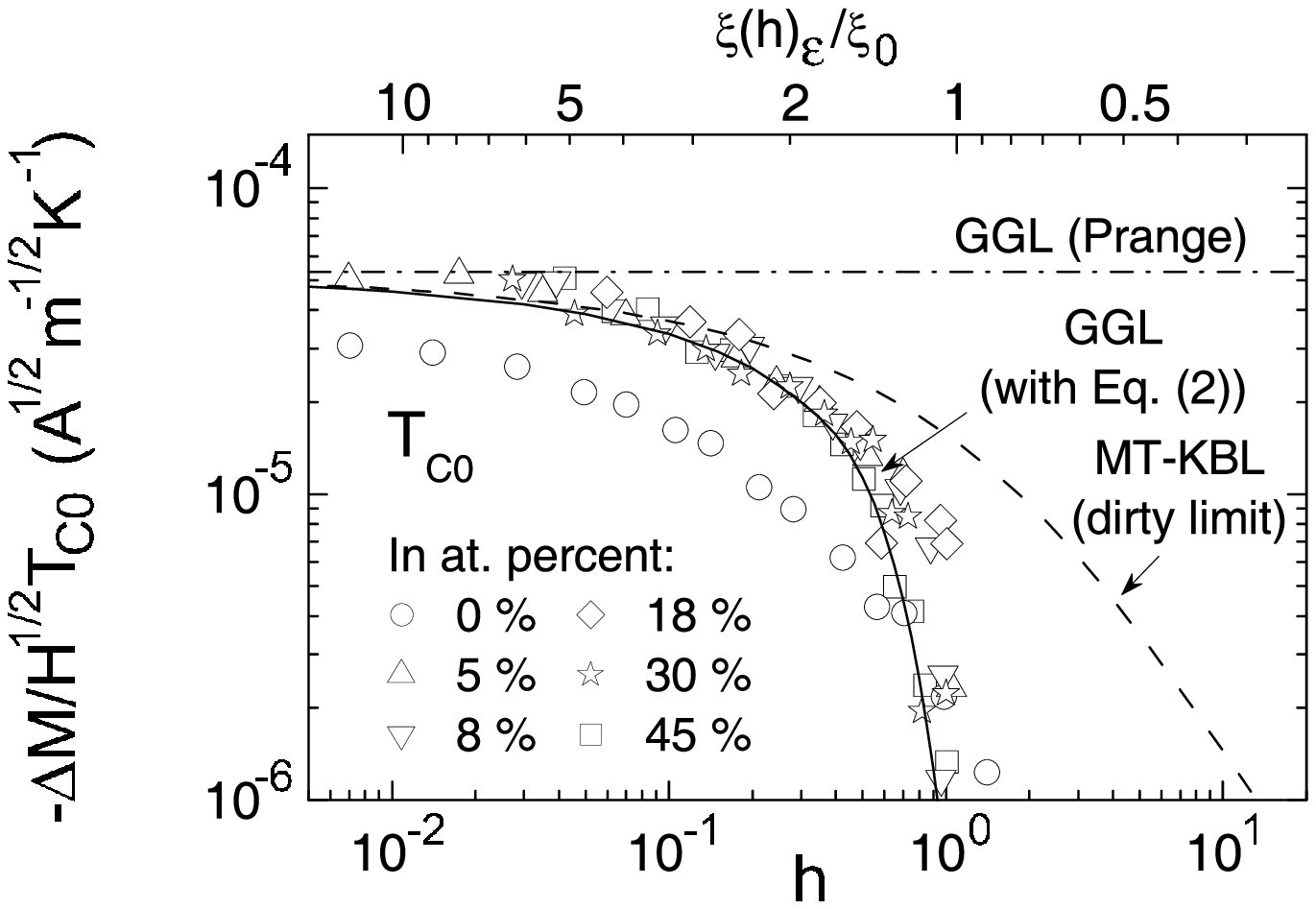}
\caption{\label{Fig3} $\Delta M(h)/H^{1/2}T_{C0}$ versus $h$ at $T_{C0}$ for all the compounds studied here. The upper scale illustrates that for all the compounds  the fluctuation effects at $T_{C0}$ vanish, sharply in this scale, when $\xi(h)_\varepsilon\simeq\xi_0$ (which corresponds to $h\simeq1.1$). The data for pure Pb are strongly affected by non-local electrodynamic effects which, even at low fields, decrease the $\Delta M$ amplitude. The lines correspond to different phenomenological and microscopic approaches, as explained in the main text.}
\end{figure}}

\newcommand{\figiv}{
\begin{figure}[t]
\epsfxsize=0.9\textwidth
\epsfbox{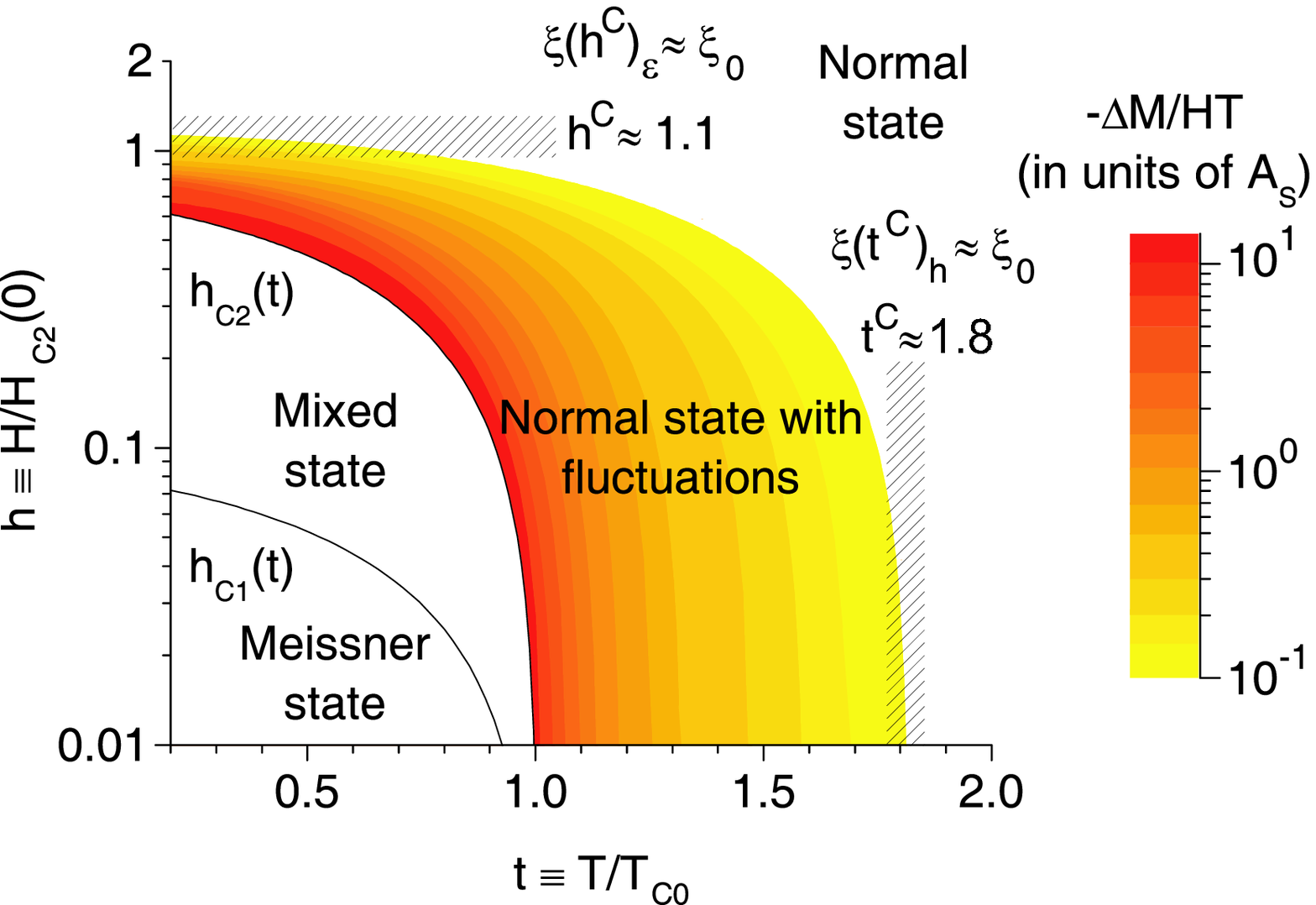}
\caption{\label{Fig4}  Measured $h-t$ phase diagram, including the SF above $h_{C2}(t)$, for the Pb-45 at.~\% In alloy. The color scale represents the fluctuation-induced magnetization (scaled by $HT$) in units of the Schmid amplitude, $A_S\equiv\pi\mu_0k_B\xi(0)/6\phi_0^2$. }
\end{figure}}

\newcommand{\tablei}{
\begin{table}[ht]
\caption{
 Main parameters of the Pb-In alloys studied in this work.   $T_{C0}$ was determined from the field-cooled $M(T)_H$  curve under a magnetic field of  0.5 mT. $H_{C2}(0)$ and the Ginzburg-Landau (GL) parameter, $\kappa$, were obtained from the reversible $M(H)_T$ curves in the mixed state. The GL coherence length amplitude, $\xi(0)$, follows from $\xi^2(0)=\phi_0/2\pi\mu_0H_{C2}(0)$. The electronic mean free path, $\ell$, was obtained from measurements of the low-temperature residual resistivity. $\xi^{Pb}_0/\ell$ is the {\it dirtiness} parameter, where $\xi^{Pb}_0\simeq 920$~\AA\ is Pippard's coherence length of pure lead [determined from $\xi_0^{Pb}=1.35\xi^{Pb}(0)$]. 
}
\begin{center}
  \mbox{}\\ 
\begin{tabular}{ccccccc}
\hline\hline
In at.~\%& $T_{C0}$ & $\mu_0H_{C2}(0)$ & $\xi(0)$ & $\kappa$ & $\ell$&$\xi^{Pb}_0/\ell$\\
& (K) & (T) & (\AA) &  & (\AA) & \\
\hline
0& 7.16 & 0.01$^a$& 680 & 0.30$^a$ & $\sim$ 50000 & $\sim$ 0\\
5& 7.06 & 0.29 & 340 & 1.3 & 200 & 2.7\\
8& 6.99 & 0.49 & 260 & 2.1 & 130 & 7.1\\
18& 6.85 & 0.85 & 200 & 3.4 & 67 & 13.7\\
30& 6.75 & 1.00 & 180 & 4.2 & 57 & 16.1\\
45& 6.43 & 1.19 & 170 & 5.5 & 42 & 21.9\\
\hline\hline &&&&&&\\
\end{tabular}
\end{center}
\mbox{}\\
{$^a$\footnotesize Extrapolated from the $H_{C2}(0)$ and $\kappa$ values of the Pb-In alloys following the Gor'kov theory.\cite{Werthamer}}
\end{table}}

\begin{document}

\title{\mbox{}\vspace{1cm}\mbox{}\\
\Large\bf
Breakdown by a magnetic field of the superconducting fluctuations in the normal state\\  in Pb$_{1-x}$In$_x$ alloys
\\ \mbox{}\\ \mbox{}\\ }

\author{
\normalsize
F\'elix Soto, Carlos Carballeira, Jes\'us Mosqueira, \\  \normalsize 
Manuel V.~Ramallo, Mauricio Ruibal, Jos\'e A.~Veira, F\'elix Vidal\\ \mbox{}\\ \normalsize
 Laboratorio de Baixas Temperaturas e Superconductividade,$^\#$\\ \normalsize
 Facultade de F\'{\i}sica, Universidade de Santiago de Compostela,
\\ \normalsize   Santiago de Compostela, E15782 Spain.
\\  \mbox{} \\ {\small \# Unidad Asociada al Instituto de Ciencias de Materiales de Madrid, CSIC, Spain.}}

\date{}
\maketitle

\begin{abstract}
The effects induced on the magnetization by coherent fluctuating Cooper pairs in the normal state have been measured in Pb$_{1-x}$In$_x$ alloys up to high magnetic fields, of amplitudes above $H_{C2}(0)$, the upper critical field extrapolated to $T$=0~K.
 Our results show that in dirty alloys these superconducting fluctuation effects are, in the entire $H-T$ phase diagram above $H_{C2}(T)$, independent of the amount of impurities and that they vanish when $H\sim1.1 H_{C2}(0)$. These striking results are consistent with a phenomenological estimate that takes into account the limits imposed by the uncertainty principle to the shrinkage, when $H$ increases, of the superconducting wave function. 
\end{abstract}

\newpage


\setlength{\textwidth}{15.5cm}
\setlength{\oddsidemargin}{0.5cm}
\setlength{\textheight}{22cm}
\setlength{\topmargin}{-1cm}
\setlength{\parskip}{18pt}

After the pioneering theoretical proposals in the sixties,\cite{ThoulessUno,AL,Smid} it was soon realized that in addition to their intrinsic interest the fluctuating Cooper pairs created in the normal state by the unavoidable thermal agitation provide a useful tool in studying the superconducting transition.\cite{ST} Since then, the superconducting fluctuations (SF) in the normal state have been extensively studied in low- and high-$T_C$ superconductors, and at present many of their main aspects are well understood.\cite{ST,Bok} However, their behaviour under strong magnetic fields [of the order of $H_{C2}(0)$, the upper critical field amplitude extrapolated to $T$=0 K] is still an open problem. In fact, a central question remains unaddressed, both experimentally and theoretically, until now: Up to what magnetic field amplitudes may the SF in the normal state survive? The interest of this question is enhanced by the fact that it may concern the general behaviour of the Cooper pairs above and below the superconducting transition in the presence of strong ``antisymmetric'' perturbations (which ``act with opposite sign on the two members of a Cooper pair''\cite{deGennes}) and it could then have implications well beyond the SF issue, including the coexistence of magnetism and superconductivity or the interplay between normal-state properties and high temperature superconductivity.\cite{Bok,Goodenough}

In this Letter,  we attempt to answer the question stated above by presenting measurements of $\Delta M(T,H)$, the magnetization induced by SF in the normal state,\cite{defDM} in Pb$_{1-x}$In$_x$ alloys with 0$\leq x\leq$0.45, and up to field amplitudes well above $H_{C2}(0)$. Our experiments show that the {\it intrinsic} SF (not affected by dynamic and non-local electrodynamic effects) are immune to the presence of non-magnetic impurities and that, independently of the superconductors' dirtiness, $\Delta M(T,H)$ vanishes when $H$ becomes close to $1.1 H_{C2}(0)$. This striking behaviour, not predicted by the existing phenomenological\cite{Smid,ST,Bok,Prange,PAW,CarballeiraPRLPhaC,MosqueiraPRL} or  microscopic\cite{LPKAE,MTMaki,Klemm} approaches for $\Delta M(T,H)$, is crudely explained here by taking into account the limits imposed by the uncertainty principle to the shrinkage of the superconducting wave function when the magnetic field increases. This will extend to high fields our previous proposals for the SF at high reduced temperatures\cite{VidalEPL}, in spite that the magnetic field is an antisymmetric perturbation.

Among the available low- and high-$T_C$ superconductors, we choose the Pb-In alloys to study the high-field behaviour of the SF for four main reasons: i)~Its entire $H-T$ phase diagram is, even for $H\gg H_{C2}(0)$, easily accessible with the existing high resolution, SQUID based, magnetometers. ii)~By changing the In concentration, it is possible to cover both type I and type II superconductors and also the range from the clean to the dirty limits (see Table I). This is a crucial advantage because it has allowed us to separate the ``universal'' magnetic field effects on the SF from those associated with the dynamic and non-local electrodynamic effects, these last being strongly material-dependent\cite{ST,MosqueiraPRL,LPKAE,MTMaki,Klemm,Gollub2,MosqueiraJPCM}. iii)~It is also possible to obtain alloys with high stoichiometric quality. This is another crucial advantage, because it minimizes the spurious magnetization rounding associated with $T_C$ inhomogeneities, that otherwise would be entangled with the intrinsic rounding due to the SF. iv) The normal-state magnetic susceptibility of these samples is almost independent of $T$ and $H$ up to, at least, $5 T_{C0}$ and $5 H_{C2}(0)$. This allows a very reliable obtainment of the background magnetization around $T_C(H)$ by linear extrapolation of the as-measured $M(T)_H$ or $M(H)_T$ well above $T_C(H)$.  Moreover, by electrochemically coating with a normal metal (i.e., Au or Cu), it is easy to eliminate in Pb-In alloys the surface superconductivity between $H_{C2}(T)$ and $H_{C3}(T)$, which otherwise would complicate the analysis above $H_{C2}(T)$ for $T<T_{C0}$, the zero-field critical temperature. 
The synthesis of this type of alloys and some of their general characteristics (including the values of the critical fields) have been already described in various earlier works which address other phenomena in these materials.\cite{FelixPb} Other experimental details will be published elsewhere.

The presence of fluctuating Cooper pairs above $T_C(H)$ produces a rounding of the as-measured susceptibility versus magnetic field curves, as illustrated in Fig.\reffig{Fig1}. This example also shows that this rounding is progressively reduced as the applied field increases and it completely vanishes when the  reduced field, $h\equiv H/H_{C2}(0)$, becomes close to 1.1. The finite-field effects may be described quantitatively through the $\Delta M(h)_\varepsilon$ curves, as the one presented in Fig.\reffig{Fig2}, which corresponds to a temperature above $T_{C0}$ [i.e., $\epsilon>0$, where $\varepsilon\equiv\ln(T/T_{C0})$ is the reduced temperature]. As may be seen in this figure, in the low-field region $\Delta M(h)_\varepsilon$ agrees with the Prange predictions\cite{Prange} and provides then a direct indication that in this dirty superconductor $\Delta M(h,\varepsilon)$ is not appreciably affected by non-local electrodynamic effects. The results of Fig.\reffig{Fig2} also show that when $h\stackrel{>}{_\sim}0.2$, $\Delta M(h)_\varepsilon$ begins to decrease and for $h\gsim1.1$ the fluctuation-induced diamagnetism vanishes. The upper horizontal scale in Fig.\reffig{Fig2}  illustrates that at these high fields the GL coherence length, which for $h\gg\vert\varepsilon\vert$ behaves as\cite{ST} $\xi(h)_\varepsilon\simeq\sqrt{2}\xi(0)h^{-1/2}$, becomes of the order of $\xi_0$, the actual (or Pippard) superconducting coherence length at $T$=0~K. This scale was obtained by using $\xi(0)=0.74\xi_0$, which is still a good approximation in the dirty limit.\cite{ST} When compared with our previous results at low field amplitudes but at high reduced temperatures\cite{CarballeiraPRLPhaC,MosqueiraPRL,VidalEPL,MosqueiraJPCM}, this last finding already suggests that in spite of the antisymmetric character of the magnetic field the vanishing of $\Delta M(h)_\varepsilon$ may also be due to the limits imposed by the uncertainty principle to the shrinkage of the superconducting wave function.\cite{VidalEPL} When the shrinkage of the superconducting wave function is due to a magnetic field, this leads to: 
\begin{equation}
\xi(h)_\varepsilon\stackrel{>}{_\sim}\xi_0,
\label{desigualdad}
\end{equation}
where $\xi_0$ for each alloy is related to $\xi_0^{Pb}$ for pure Pb by\cite{ST} $\xi_0\simeq(\xi_0^{Pb}\ell)^{1/2}$. The above inequality directly leads to a critical reduced-field, $h^C$, given by $h^C=2(\xi(0)/\xi_0)^2$, above which all the SF vanish. By using again $\xi(0)=0.74\xi_0$, we obtain $h^C\simeq1.1$, in excellent agreement with the results of Figs.\reffig{Fig1} and\reffig{Fig2}. As $\xi(0)/\xi_0$ is almost material-independent,\cite{deGennes,Werthamer} Eq.~(1) predicts that the above value of $h^C$ will be ``universal''. This striking prediction is confirmed at a quantitative level by the experimental results at $T_{C0}$ for all the samples studied in this work and summarized in Fig.\reffig{Fig3}: Independently of the superconductor dirtiness and also of their type I or type II character, the SF vanish when $h\simeq1.1$. Another central result shown by Fig.\reffig{Fig3} is that all the data for the different dirty alloys collapse on the same curve: this provides a direct experimental demonstration that in the absence of non-local electrodynamic effects, the SF are not affected by impurities.

The data summarized in Fig.\reffig{Fig3} are also particularly well adapted to probe in the high-field region below $h^C$ the applicability of the existing theoretical approaches for $\Delta M(h)_\varepsilon$: in the absence of material-dependent effects (as dynamic or non-local effects), the Gaussian-Ginzburg-Landau (GGL) approach without any cutoff (as first proposed by Prange\cite{Prange}) predicts that at $T_{C0}$ all the data must collapse on the same, field-independent curve (dot-dashed line in Fig.\reffig{Fig3}). The failure of the Prange approach when $h$ increases, also shown by the results of Fig.\reffig{Fig2}, was already observed by Gollub and coworkers in their pioneering measurements\cite{ST,Gollub2}, which did not cover the high-field regime (they extended only up to $h\sim 0.6$). Most of these last measurements were done in clean low-$T_C$ superconductors and they were entangled with the non-local effects that already at low fields reduce the $\Delta M(h)_{T_{C0}}/H^{1/2}T_{C0}$ amplitude well below the theoretical value. We have also observed this reduction in pure Pb, as illustrated in Fig.\reffig{Fig3}. The introduction in the GGL Prange approach of different cutoffs, which consider the short-wavelength fluctuation effects at high reduced temperatures and fields\cite{PAW,CarballeiraPRLPhaC,MosqueiraPRL} but do not take into account Eq.~(1) (see also below), will lead to a decrease of $\Delta M(h)_\varepsilon$ when $h$ increases. However, as illustrated in Fig.\reffig{Fig2} for the case of the kinetic-energy or momentum cutoff\cite{CarballeiraPRLPhaC}, the resulting $\Delta M(h)_\varepsilon$ (dotted line) still does not account for the experimental decrease with $h\gsim0.2$.

The results in Figs.\reffig{Fig2} and\reffig{Fig3} also show that the existing microscopic approaches for $\Delta M$\cite{LPKAE,MTMaki,Klemm}, which do not take into account Eq.~(1), also fail to account for our experimental results at high fields. The dashed lines correspond to the approach proposed by Maki and Takayama\cite{MTMaki} and by Klemm, Beasley and Luther\cite{Klemm} (MT-KBL theory) for the dirty limit. This approach generalizes for dirty superconductors the pioneering calculations of Lee and Payne.\cite{LPKAE} We have estimated these curves without any adjustable parameter (by using the values of Table I). As may be seen in Figs.\reffig{Fig2} and\reffig{Fig3}, the agreement with the experimental data at low and moderate field amplitudes (up to $h\simeq0.2$) is excellent. However, there is strong disagreement at high fields.

Could the theoretical approaches summarized above explain the high-field behaviour of $\Delta M$ if Eq.~(1) is taken into account? A full answer to this question could be obtained only on the grounds of the microscopic approaches \cite{LPKAE,MTMaki,Klemm,Tesanovic}. This task is out of the scope of our present Letter. However, a crude but probably the simplest way to obtain a qualitative answer to this question is to introduce Eq.~(1) in the GGL approach, even though far from $T_{C0}$ or at high fields this approach may be not formally applicable.\cite{ST,VidalEPL} In fact, in this way we may probe if the limitations imposed by the uncertainty principle is the dominant effect on the SF at high magnetic fields, in spite of the antisymmetric character of the magnetic field. To do that, we first note that in terms of the ``total energy'' $E_{nk_z}$ of the fluctuating modes of Landau level index $n=0,1,...$ and wave vector parallel to the field $k_z$, this constraint may be written as:
\begin{equation}
E_{nk_z}\equiv\epsilon+(2n+1)h+\xi^2(0)k_z^2\;\lsim\;(\xi(0)/\xi_0)^2,
\label{cutH}
\end{equation}
where the energies are expressed in units of  $\hbar^2/2m^*\xi^2(0)$, and $\hbar$ and $m^*$ are, respectively, the Planck constant and the effective mass of the Cooper pairs. To take into account Eq.~(2) at high fields, we introduce a total-energy-dependent weighting function, \wEnk , pondering the contribution of each fluctuating mode in the free-energy statistical sum. This procedure is similar to the one proposed by Patton, Ambegaokar and Wilkins (PAW).\cite{PAW} However, PAW's approach does not take into account the limits imposed by the uncertainty principle to the shrinkage of the superconducting wave function. In fact, PAW's calculations are equivalent to the choice $\wEnkPAW=\ln[1+\exp(-E_{nkz}/h_0)]/\ln E_{nk_z}$, which does not consider the inequality (2). Here the reduced field $h_0$ corresponds to the maximum of the $\Delta M(h)_{T_{C0}}$ curve, and therefore in our Pb-In alloys it will be $0.2\lsim h_0\lsim0.25$. In our present study, to reproduce the rapid fall-off of the SF expected when the inequality (2) begins to be violated, we introduce an additional prefactor to the penalization function, using $W(E_{nk_z})=W_{\rm PAW}(E_{nk_z})(1+\exp[(E_{nk_z}-(\xi(0)/\xi_0)^2-\delta)/\delta])^{-1}$. This additional prefactor has the form of a Fermi-Dirac distribution function, presenting a step-like decay starting at energies $\sim(\xi(0)/\xi_0)^2$ and with $\delta$ as typical half-width. By repeating the standard GGL calculations for \DMTH\ in isotropic 3D superconductors above the transition\cite{Prange,PAW,CarballeiraPRLPhaC,MosqueiraPRL} but now including the weighting function \wEnk, we obtain:
\begin{equation}
\DM=
{\frac{k_{\rm B} T} {\pi\phi_0}}\hskip-0.5em
\int_{0}^{\infty}\hskip-1em{\rm d}k_z\sum_{n=0}^\infty{\frac{\partial}{\partial h}}\left[
\wEnk h \ln E_{nk_z}\right],
\label{Mfl}
\end{equation}
where $k_B$ is the Boltzmann constant and $\phi_0$ the flux quantum. This formula may be numerically computed thanks to the rapid decay of $W(E_{nk_z})$ as $n$ or $k_z$ increase.  In Figs.\reffig{Fig2} and\reffig{Fig3} we plot the results of that evaluation as a solid line. In making these computations we have used the values of $\xi(0)$ and $T_{C0}$ given in Table I and $(\xi(0)/\xi_0)^2=0.55$. We also used $h_0=0.22$ and $\delta=0.2$, which are the values giving a better agreement with our experimental results. As may be seen in Figs.\reffig{Fig2} and\reffig{Fig3}, this agreement is excellent in all the studied $h$- and $\varepsilon$-range, and it also includes the vanishing of $\Delta M(h)_\varepsilon$ at $h^C\simeq1.1$.

Another well-known procedure to shrink the collective wave function of the fluctuating Cooper pairs is, as noted before, to increase the temperature well above $T_{C0}$ in the absence of or under low magnetic fields. In fact, we have recently studied in detail the behaviour of the SF in this case in different low-$T_C$ and high-$T_C$ superconductors through measurements of the paraconductivity or of $\Delta M(\varepsilon)_h$ in the low field limit ($h/\varepsilon\ll 1$)\cite{MosqueiraPRL,VidalEPL,MosqueiraJPCM,MosqueiraEPLCarballeiraDs,CurrasPRBVinaPRB}. When both types of experiments are compared, they further suggest a similar pair breaking mechanism when $\varepsilon$ approaches $\varepsilon^C\simeq0.55$ or $h$ approaches $h^C\simeq1.1$, despite the magnetic field being an ``antisymmetric'' perturbation\cite{deGennes}. These results also demonstrate that in the absence of non-local electrodynamic effects (which affect only the clean and low-$\kappa$ Pb), the SF in the normal state are unaffected by impurities (as sown in Fig.\reffig{Fig3}). This immunity of the SF against both impurities and (antisymmetric) magnetic field perturbations, which is also confirmed by measurements of the paraconductivity at high reduced temperatures in high-temperature cuprate superconductors (HTSC) with different doping levels\cite{VidalEPL,CurrasPRBVinaPRB}, is to some extent similar to the stability shown by the flux quantization below $H_{C2}(T)$. Our results suggest then the existence of an unexpected ``quantum protectorate''\cite{LaughlinAnderson} for the coherent fluctuating Cooper pairs above $H_{C2}(T)$, that is only broken by the limits imposed by the uncertainty principle. Its extension in the $h-t$ phase diagram is represented in Fig.\reffig{Fig4}, which corresponds to the Pb-45 at.~\% In alloy.

In conclusion, we believe that the {\it measured} ($h$-$t$) phase diagram shown in Fig.\reffig{Fig4} is representative of all type II superconductors unaffected by non-local electrodynamic effects, as the dirty low-$T_C$ superconductors. In principle, this ($h$-$t$) phase diagram could also apply to HTSC, which are extremely type II superconductors also unaffected by non-local effects. However, this last suggestion obviously needs further experimental verification, in particular in view of the recent observation of anomalous thermomagnetic effects well above $T_C(H)$ in some underdoped cuprates.\cite{termomagneticos} Other open questions which deserve further studies are the SF in presence of magnetic order (our present results also suggest the robustness of the SF against this antisymmetric perturbation) or the relationships between our crude theoretical analysis and the available microscopic approaches for the SF around $T_C(H)$ \cite{LPKAE,MTMaki,Klemm,Tesanovic}.

\mbox{}

\mbox{}

We thank Prof.~J.B.~Goodenough for his careful reading of our manuscript and helpful remarks. We acknowledge the financial support from the ESF ``Vortex'' Program,  the CICYT, Spain, under grants no.~MAT2001-3272 and MAT2001-3053,  the Xunta de Galicia under grant PGIDT01PXI20609PR, and by Uni\'on Fenosa under grant 220/0085-2002.

\pagebreak

\pagebreak

\mbox{}\vspace{2cm}\mbox{}\\

\tablei

\figi

\figii

\figiii

\figiv


\begin{thebibliography}{99}
\bibitem{defDM} The fluctuation-induced magnetization is defined as $\Delta M(T,H)\equiv M(T,H)-M_B(T,H)$, where $M(T,H)$ is the as-measured magnetization and $M_B(T,H)$ the so-called background or bare magnetization [which may be approximated by extrapolating the magnetization measured either at $T\gg T_{C0}$ or $H\gg H_{C2}(0)$].
\end{thebibliography}
\end{document}